\documentclass{aa} %[referee]
\usepackage{txfonts}
\usepackage{graphicx, graphics}
\usepackage{CJK}
\usepackage{bbm}
\usepackage{amsmath, amssymb, bm}
\usepackage[nodayofweek]{datetime}
\input{prepcitationformat.txt}
\usepackage{txfonts}
\usepackage{changes}

\usepackage{tikz}

\definecolor{lime}{HTML}{A6CE39}
\DeclareRobustCommand{\orcidicon}{%
    \raisebox{-3pt}{\begin{tikzpicture}
    \filldraw [lime, yshift=-2pt] (0, 0) circle [radius=0.16]
    node[white] {\raisebox{1pt}{\hspace{0.5pt}\fontfamily{qag}\selectfont\tiny i\scalebox{0.8}{D}}};
    \end{tikzpicture}}
    \hspace{-2.5mm}
    \vspace{-0.25pt}
}

\global\def\tablenotemark#1{{\color{blue}{\normalfont\textsuperscript{\scriptsize #1}}}} % changed by Bin Ren to \scriptsize to match published tablenotemark

\newdateformat{monthyearday}{% format preparation to display today in DD Month YYYY format
  \THEYEAR\ \monthname[\THEMONTH] \twodigit{\THEDAY}}
\newdateformat{daymonthyear}{%
  \twodigit{\THEDAY}\ \monthname[\THEMONTH] \THEYEAR}

\newcommand{\orcidauthor}[2]{#2\href{http://orcid.org/#1}{\orcidicon}}

\titlerunning{Vega in M-band}
\authorrunning{Ren et al.}

\begin{document}
\begin{CJK*}{UTF8}{gbsn}
\title{Planet Search with the Keck/NIRC2 Vortex Coronagraph in $M_s$-band for Vega}

\author{
\orcidauthor{0000-0003-1698-9696}{Bin B. Ren (任彬)}\inst{\ref{inst-oca}, \ref{inst-uga}, \ref{inst-cit}}
\and
\orcidauthor{0000-0003-0354-0187}{Nicole L. Wallack}\inst{\ref{carnegie}, \ref{inst-cit-gps}} 
\and
\orcidauthor{0000-0002-6903-9080}{Spencer A. Hurt} \inst{\ref{inst-oregon}}
\and
\orcidauthor{0000-0002-8895-4735}{Dimitri Mawet}\inst{\ref{inst-cit}, \ref{inst-jpl}}
\and
\orcidauthor{0000-0001-5365-4815}{Aarynn L. Carter} \inst{\ref{inst-ucsc}}
\and
\orcidauthor{0000-0002-1583-2040}{Daniel Echeverri} \inst{\ref{inst-cit}}
\and
\orcidauthor{0000-0002-3414-784X}{Jorge Llop-Sayson}\inst{\ref{inst-cit}}
\and
\orcidauthor{0000-0001-6126-2467}{Tiffany Meshkat} \inst{\ref{inst-ipac}}
\and
\orcidauthor{0000-0001-7130-7681}{Rebecca Oppenheimer}  \inst{\ref{inst-amnh}}
\and
\orcidauthor{0000-0003-3184-0873}{Jonathan Aguilar} \inst{\ref{inst-stsci}}
\and
Eric Cady \inst{\ref{inst-jpl}}
\and
\orcidauthor{0000-0002-9173-0740}{\'Elodie Choquet}  \inst{\ref{inst-lam}} % C314 observer
\and
\orcidauthor{0000-0003-4769-1665}{Garreth Ruane} \inst{\ref{inst-jpl}}
\and
\orcidauthor{0000-0002-1871-6264}{Gautam Vasisht}  \inst{\ref{inst-jpl}}
\and
\orcidauthor{0000-0001-7591-2731}{Marie Ygouf}  \inst{\ref{inst-jpl}}% C314 observer
}

\institute{
Universit\'{e} C\^{o}te d'Azur, Observatoire de la C\^{o}te d'Azur, CNRS, Laboratoire Lagrange, F-06304 Nice, France; \email{\url{bin.ren@oca.eu}
% \url{bren2@alumni.jh.edu}
} \label{inst-oca}
\and
Univ. Grenoble Alpes, CNRS, IPAG, F-38000 Grenoble, France \label{inst-uga}
\and
Department of Astronomy, California Institute of Technology, MC 249-17, 1200 East California Boulevard, Pasadena, CA 91125, USA \label{inst-cit}
\and
Earth and Planets Laboratory, Carnegie Institution for Science, Washington, DC 20015, USA\label{carnegie}
\and
Division of Geological \& Planetary Sciences, California Institute of Technology, MC 150-21, 1200 East California Boulevard, Pasadena, CA 91125, USA \label{inst-cit-gps}
\and
Department of Earth Sciences, University of Oregon, Eugene, OR 97403, USA \label{inst-oregon}
\and
Jet Propulsion Laboratory, California Institute of Technology, 4800 Oak Grove Drive, Pasadena, CA 91109, USA \label{inst-jpl}
\and
Department of Astronomy \& Astrophysics, University of California, Santa Cruz, 1156 High St, Santa Cruz, CA 95064, USA \label{inst-ucsc}
\and
Infrared Processing and Analysis Center (IPAC), California Institute of Technology, MC 100-22, 1200 East California Boulevard, Pasadena, CA 91125, USA \label{inst-ipac}
\and
American Museum of Natural History, New York, NY 12345, USA \label{inst-amnh}
\and
Space Telescope Science Institute (STScI), 3700 San Martin Drive, Baltimore, MD 21218, USA \label{inst-stsci}
\and
Aix Marseille Univ, CNRS, CNES, LAM, Marseille, France \label{inst-lam}
}

\date{Received 12 July 2022 / Revised 17 January 2023 / Accepted 18 January 2023} % \daymonthyear\today\ % show today's date in DD Month YYYY format

\abstract
{Gaps in circumstellar disks can signal the existence of planetary perturbers, making such systems preferred targets for direct imaging observations of exoplanets.}
{Being one of the brightest and closest stars to the Sun, the photometric standard star Vega hosts a two-belt debris disk structure. Together with the fact that its planetary system is being viewed nearly face-on, Vega has been one of the prime targets for planet imaging efforts.}
{Using the vector vortex coronagraph on Keck/NIRC2 in $M_s$-band at $4.67~\mu$m, we report the planet detection limits from $1$~au to $22$~au for Vega with an on-target time of $1.8$~h.}
{We reach a $3~M_{\rm Jupiter}$ limit exterior to $12$~au, which is nearly an order of magnitude deeper than existing studies. Combining with existing radial velocity studies, we can confidently rule out the existence of companions more than ${\sim}8~M_{\rm Jupiter}$ from $22$~au down to $0.1$~au for Vega. Interior and exterior to ${\sim}4$~au, this combined approach reaches planet detection limits down to ${\sim}2$--$3~M_{\rm Jupiter}$ using radial velocity and direct imaging, respectively.}
{By reaching multi-Jupiter mass detection limits, our results are expected to be complemented by the planet imaging of Vega in the upcoming observations using the \textit{James Webb Space Telescope} to obtain a more holistic understanding of the planetary system configuration around Vega.}

\keywords{stars: imaging -- planets and satellites: detection -- techniques: high angular resolution -- techniques: image processing -- planets and satellites: individual: Vega}

\maketitle

\section{Introduction}
Vega ($\alpha$ Lyrae,  \textit{Keoe}, \textit{w\=aqi'}, \textit{Zhin\"{u}} 织女星), one of the historical photometric standard stars \citep{Johnson53}, is an A0~V star \citep[e.g.,][]{Johnson53} that is located at $7.68\pm0.02$~pc from the Solar System \citep{vanleeuwen07}. Despite its proximity, Vega is not included in the \textit{Gaia} Catalog of Nearby Stars that are within $100$~pc from the Sun, since its brightness exceeds the \textit{Gaia} limits \citep{GCNS21}. With an age of $445\pm13$~Myr \citep{yoon10}, Vega shows a prototypical mid-infrared excess in the \textit{IRAS} observations, situating it in an evolutionary stage that is between star formation and our Solar System \citep{aumann84}.

The two-belt debris disk system around Vega may result from planet-disk interaction under various planetary configurations \citep[e.g.,][]{matra20}. In fact, most existing directly imaged planets are found in bright debris disk systems (e.g., HR~8799: \citealp{marois08}, $\beta$~Pic: \citealp{langrange09}, 51 Eri: \citealp{macintosh15}), making it more likely to find giant planets in debris disk systems than around stars without disks \citep{Meshkat17}. Combined with the observational fact that giant planets are more likely to exist at $1$--$10$~au from their host stars in both direct imaging and radial velocity surveys \citep{nielsen19, fulton21}, the proximity of Vega makes it one of the best systems for giant planet search.

Before the resolved imaging of debris belts around Vega, dust structures around Vega have suggested the existence of potential planetary perturbers \citep[e.g.,][]{holland98, wilner02} and called for deep imaging of them. Combining the Gemini Altar adaptive optics (AO) system and the NIRI instrument, \citet{marois06} obtain $5\sigma$ detection limits of better than 18 Mag at $3\arcsec$--$10\arcsec$ in the off-methane $1.58\mu$m $6.5\%$ filter, or ${\sim}3~M_{\rm Jupiter}$ at $8\arcsec$. Using the MMT AO and the Clio camera, \citet{hinz06} and \citet{heinze08} obtain $10\sigma$ limits of better than ${\sim}12$~Mag at $2\arcsec$--$11\arcsec$ at $M$-band,  or ${\sim}10~M_{\rm Jupiter}$.

Recent observational studies on the Vega planetary system are in direct imaging, transiting, radial velocity, and (sub)millimeter interferometric imaging. In direct imaging, \citet{meshkat18} presented observations from the coronagraphic integral field spectrograph P1640 at Palomar Observatory in $J$ and $H$ bands. Despite a non-detection of planets, they obtained $5\sigma$ planet detection limits from $0\farcs25$ to $2\arcsec$, reaching a best sensitivity of ${\approx}20~M_{\rm Jupiter}$ at $1\farcs5$ or $12$~au. From a complimentary approach, using the Atacama Large Millimeter/submillimeter Array (ALMA), \citet{matra20} observed and modeled the outer belt from ${\sim}60$~au to ${\sim}200$~au. To explain the observed disk architecture, \citet{matra20} discussed three mechanisms which include a single giant planet, multiple low-mass planets, and no outer planets. Combining a decade of TRES spectra for radial velocity and two sectors of \textit{TESS} photometry, and under the scenario that the planetary orbits have inclinations between $1\fdg5$ and $11\fdg5$, \citet{hurt21} obtained non-detection of $1$--$10~M_{\rm Jupiter}$  planets within $1$--$10$~au, while reporting a candidate Jovian signal with a period of $2.43$ day in radial velocity measurements. Nevertheless, depending on the orbit orientations of the planets, the mass limit can vary by up to a factor of ${\sim}10$ in Fig.~6 of \citet{hurt21}. 

To improve existing high-contrast imaging limits, test different mechanisms for the formation of Vega debris disk system observed in ALMA, and explore beyond the nearly edge-on limitations from radial velocity and transit studies, we observed Vega in $M_s$-band using the vortex coronagraph on Keck/NIRC2. In $M_s$-band, planets have relatively larger planet-to-star brightness ratios than in $J$- or $H$-band \citep[e.g.,][]{spiegel12, skemer14}, and the advantage of $M_s$-band over shorter-wavelength bands improves for intermediate system ages for a planet of a given mass \citep[e.g.,][]{currie22}, both enabling us to explore around Vega for possible cooler and less massive planets than existing studies.

\section{Observation and Data Reduction}\label{sec-obs}
We observed Vega using the Keck/NIRC2 vortex coronagraph in $M_s$-band in two individual nights using the narrow camera with a pixel size of $9.942$~mas (e.g., \citealp{service16}; \citealp{mawet19}). The first observation is on UT 2018 August 30 %\footnote{\url{https://www2.keck.hawaii.edu/observing/keckSchedule/keckSchedule.php?date=2018-08-29}} 
under program C314 (PI: D.~Mawet), the total integration time is $1991$~s ($=0.181$~s $\times100$~coadds $\times110$~frames), and the parallactic angle change is $74\fdg9$. The second observation is on UT 2019 August 20 %\footnote{\url{https://www2.keck.hawaii.edu/observing/keckSchedule/keckSchedule.php?date=2019-08-19}} 
under program N097 (PI: T.~Meshkat), the total integration time is $4500$~s ($=0.25$~s $\times150$~coadds $\times120$~frames), and the parallactic angle change is $81\fdg6$. We list the observation details in Table~\ref{tab-obs}.

\begin{table}
\caption{Keck/NIRC2 vortex coronagraph observation log\label{tab-obs}}
\begin{tabular}{c r r} \hline\hline
Target 					& \multicolumn{2}{c}{Vega} 		\\
Filter 					& \multicolumn{2}{c}{$M_s$-band} \\ \hline
UT Date 					&  	2018 Aug 30 	& 2019 Aug 20 \\
UT Start 					&	05:15:26.35	&	06:35:47.62	\\ 
UT End					&	07:15:00.52	& 	10:12:23.95	\\
Airmass$\tablenotemark{a}$	& $1.066\pm0.009$	&	$1.159\pm0.101$ \\
DIMM Seeing$\tablenotemark{a}^,\tablenotemark{b}$				& 	$0\farcs42\pm0\farcs11$		&	$1\farcs18\pm0\farcs55$ \\
MASS Seeing$\tablenotemark{a}^,\tablenotemark{b}$				& 	$0\farcs16\pm0\farcs06$		&	$0\farcs18\pm0\farcs08 $	\\
Precipitable Water Vapor level	& ${\sim}7$~mm	&	${\sim}2$~mm \\
Parallactic Angle Change 		& 	$74\fdg9$		&	$81\fdg6$	\\ \hline
Single Integration Time		& 	$0.181$ s 	 	& 	$0.25$ s 	\\
Coadd For Single Frame		& 	$100$		& 	$150$  	\\
Total Frame Count 			& 	$110$ 		& 	$120$  	\\
On-target Integration Time	& 	$1991$ s	 	&	$4500$ s	\\
Total On-target Time			& \multicolumn{2}{c}{$1.803$ h} 	\\ 
Pixel Scale				& \multicolumn{2}{c}{$9.942$~mas} 	\\ \hline
\end{tabular}
\begin{flushleft}
\tiny  \textbf{Notes}: $^a${The uncertainties in this Letter are $1\sigma$ unless otherwise specified.}
$^b${Calculated during this observation using \url{http://mkwc.ifa.hawaii.edu/current/seeing/index.cgi}.}
%}
\end{flushleft}
\end{table}

Faint planetary signals are often overwhelmed by the stellar point spread function (PSF) of the central star. In order to remove the stellar PSF and reveal faint surrounding signals, we first preprocess the data using the {\tt VIP} package \citep{GomezGonzalez2017}, which is further customized for NIRC2 vortex observations by performing flat-fielding, bad pixel and background removal, and image centering \citep{xuan18}. We then reduce the preprocessed data using the Karhunen--Lo\`eve image projection algorithm \citep[KLIP;][]{soummer12, amara12} that performs principal component analysis to capture the stellar PSF and speckles in the observation. For an image, we remove its stellar PSF and speckles by first projecting it to the KLIP components, and subtract the projection from the original image to obtain the residuals. Astrophysical signals including planets and disks will then reside in these residuals. To obtain the final image for each observation, we obtain the residual maps for each individual readout, then rotate and median combine them.

\begin{figure}[htb!]
%\centering
	\includegraphics[width=0.52\textwidth]{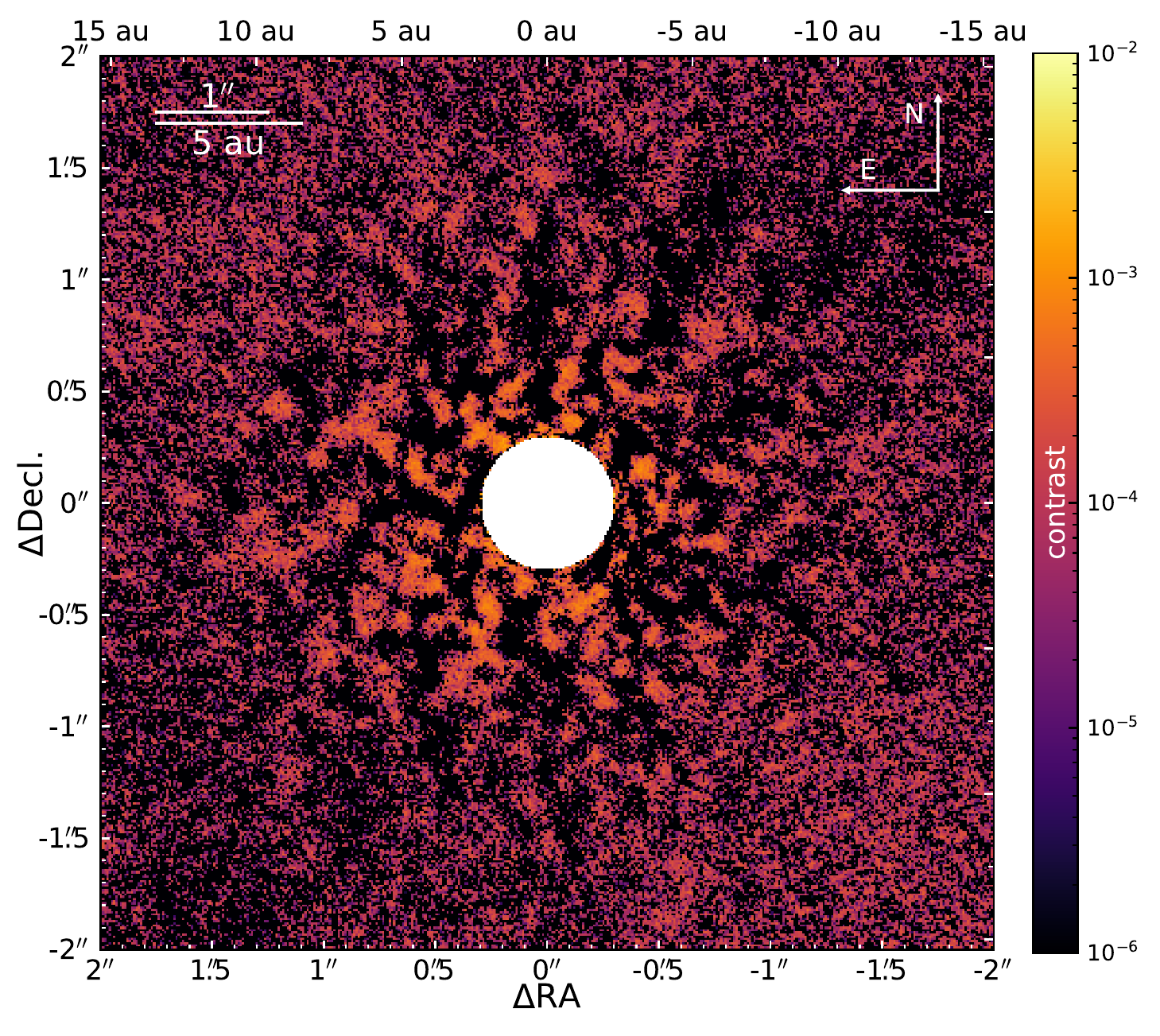}
    \caption{Combined two-epoch NIRC2 image of Vega in $M_s$-band using ADI with $10\%$ of the KLIP components for demonstration purposes. We do not identify point sources that are of more than $5\sigma$ levels beyond the noise of similar angular separation from the star. The pixel values correspond to lower limits of contrast values due to over- and self-subtraction with KLIP and ADI, respectively.}     \label{fig-image}    
\end{figure}

We present the combined two-epoch result using $10\%$ of the KLIP components with angular differential imaging (ADI: \citealp{marois06}) in Fig.~\ref{fig-image} for demonstration purposes, in which we do not identify point sources that are beyond $5\sigma$ of the noise that is within the same angular separation from the stars. 

Using \texttt{VIP} and taking into account of self-subtraction and over-fitting with ADI while using KLIP, we generate the $5\sigma$ contrast limits for each observation following \cite{xuan18} by varying the reduction parameters to obtain the corresponding detection limits while performing corrections for small sample statistics in \citet{mawet14}. Specifically, {\tt VIP} performs injection recovery for companions at different locations to measure the throughput from ADI and KLIP \citep{GomezGonzalez2017}. In measuring our throughput, we inject companions along three radial branches spread throughout the image (originating from the masked center) where the averaged throughput at each radial location in the image is determined from these multiple estimates of the throughput at different branches \citep{xuan18}. We compute the contrast for the entire image for each combination of inner and outer mask size and number of principal components, where we compute up to 30 principal components.  The ADI reduction is performed with no rotation gap; for one combination of inner and outer radii from Table~1 of \citet{xuan18}, the full-frame reduction is performed on the an annulus zone with the region interior to the outer radius or exterior to the inner radius included (e.g., for an outer radius of $0\farcs5$ and an inner radius of $0\farcs08$, all pixels with radial separations between $8$~pixel and $50$ pixel from the center of the image, or between $0\farcs08$ and $0\farcs5$, are included in the reduction; see the reduction details in \citealp{xuan18}). While we do not use annular ADI, we do utilize the best contrast achieved from our full-frame ADI at each one pixel annulus.

To obtain the final detection limit, for each angular separation from the star with a step size of 1 pixel, we compare the detection limits from different combinations of reduction parameters. The reduction parameters including frame size (i.e., algorithmic inner and outer radii) and the number of principal components, see Sect.~2 and Table~1 of \citet{xuan18} for the details on how on computing most optimal contrasts using five different frame size combinations. Therefore, while our optimal contrast is the combination of frames processed using different parameters, we are optimizing at each radial location. For our observations, the number of principal components adopted for the final contrast curve ranges from 8 to 29, with a median of 15 and a standard deviation of 7.4.

\begin{figure*}[htb!]
%\centering
	\includegraphics[height=8.25cm]{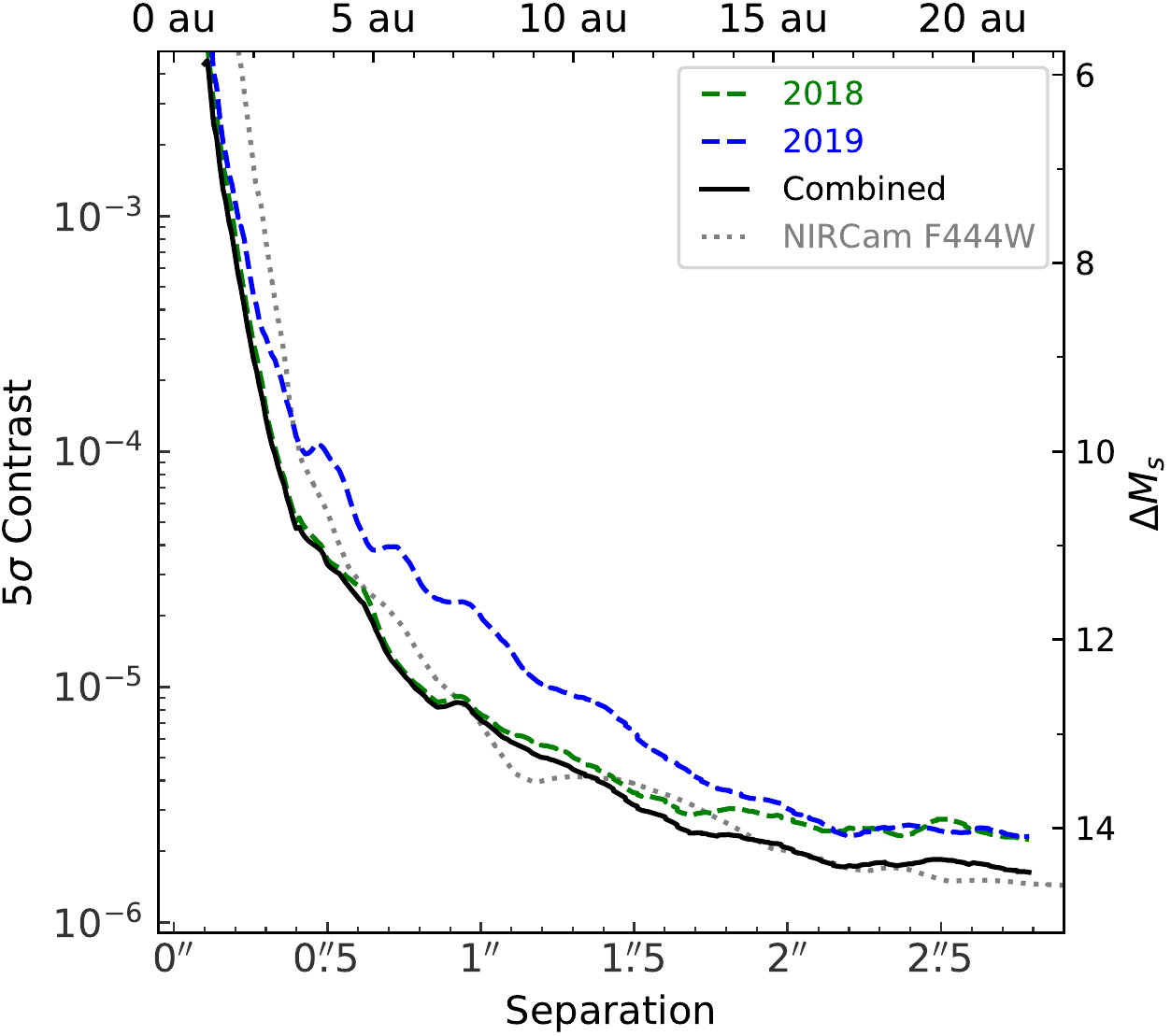}
	\includegraphics[height=8.25cm]{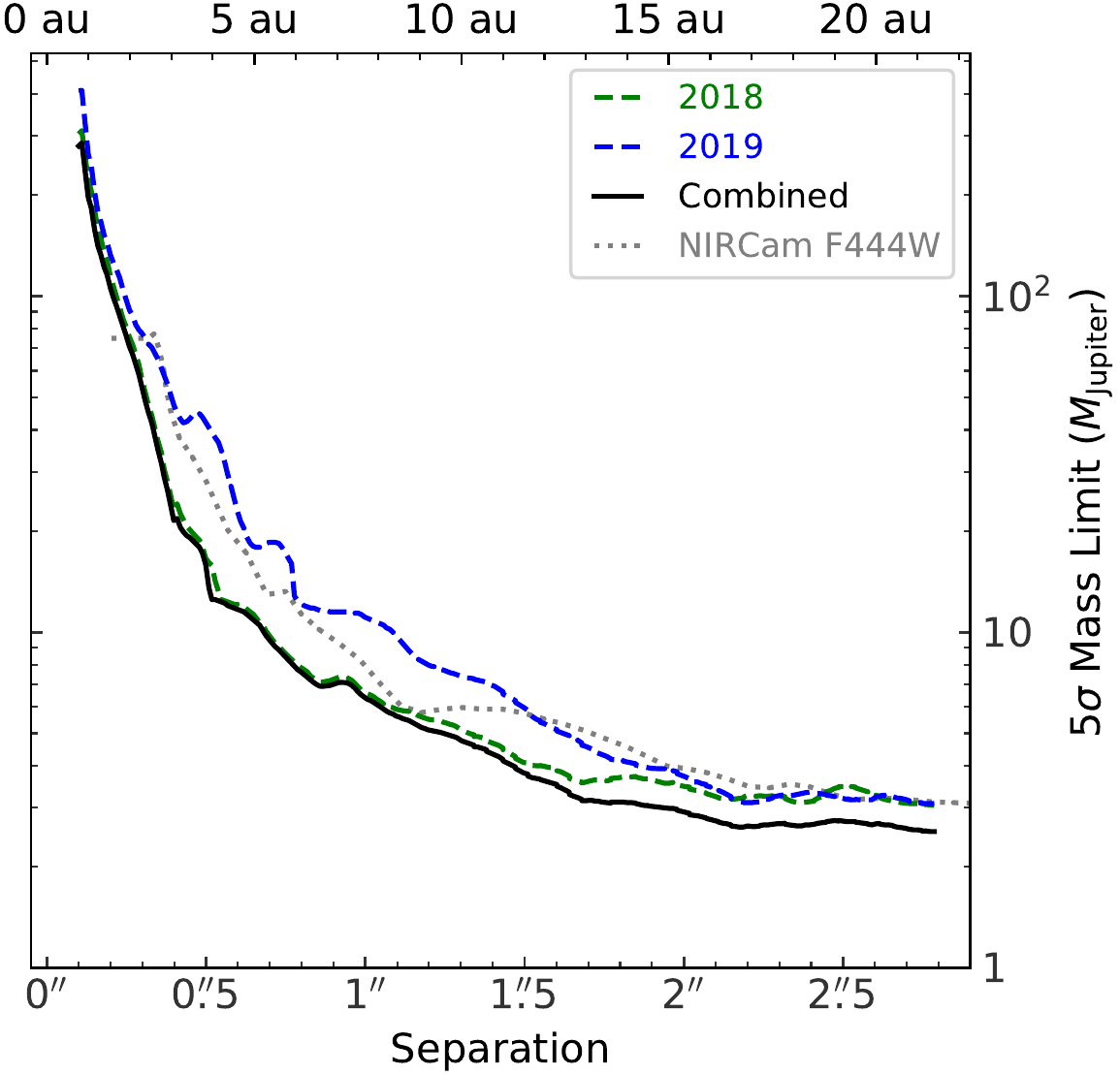}
    \caption{Detection limits of point sources around Vega in $M_s$-band. Left: $5\sigma$ contrast, in comparison with \textit{JWST}/NIRCam F444W on-sky contrast calculated from \citet{carter22}. Right: $5\sigma$ mass detection limit, and expected mass detection limit with NIRCam F444W. Note: the NIRCam F444W values have assumed identical instrument performance in the \textit{JWST}/ERS-1386 program, see Sect.~\ref{sec-jwst-sim}.}     \label{fig-contrast}    
\end{figure*}

The faintest companion that can be detected from such combinations at $5\sigma$ level is adopted as our final detection limit. For the 2019 data, the exposures that were used to image the unblocked central source were saturated, we thus fit the unsaturated first Airy ring to that of a theoretical model of the vortex stellar PSF (while taking into account of the PSF broadening effects due to weather by convolving a 2-dimensional Gaussian distribution), and use the best-fit model to generate the corresponding contrast curve. We then combine our detection limits from the two observations, and present them in Fig.~\ref{fig-contrast}. Although the 2018 observation has a shorter total on-target exposure time than the 2019 one, their total parallactic angle change difference is only $6\fdg7$. With the DIMM seeing of $0\farcs41\pm0\farcs11$ in 2018 being more stable than that of $1\farcs18\pm0\farcs55$ in 2019, the 2018 data dominates the detection limits in the combined dataset.

\section{Analysis}\label{sec-ana}

\subsection{Mass detection limits}
\subsubsection{NIRC2 imaging}

Being the photometric standard, Vega's apparent magnitude is defined as $0$ in $M_s$-band. Adopting a distance of $7.68\pm0.02$~pc in \citet{vanleeuwen07}, the absolute magnitude for Vega in $M_s$-band is $0.573 \pm 0.006$.

With an age of $445\pm13$~Myr \citep{yoon10} and adopting the AMES-Cond evolutionary models \citep{baraffe2003}, we convert the contrast to $5\sigma$ mass detection limits in Fig.~\ref{fig-contrast}. We reach a detection limit of less than $5~M_{\rm Jupiter}$ beyond $9$~au, and $3~M_{\rm Jupiter}$ beyond $14$~au.

\subsubsection{NIRC2 imaging and TRES radial velocity}\label{sec-comb-nirc2-tres}
Combining the 2018 NIRC2 results with the radial velocity data from the Tillinghast Refector Echelle Spectrograph (TRES) in \citet{hurt21}, we follow \citet{hurt21} to obtain the mass limits assuming the planetary orbits are well-aligned with the spin axis of Vega.

To explore the detectability of companions from both direct imaging and radial velocity measurements, we randomly generate $10^6$ radial velocity samples of companion orbits following \citet{hurt21}.  The semi-major axis follows a log-uniform distribution ranging from $0.1$~au to $22$~au. The companion mass follows a log-uniform distribution ranging from $0.1$ to $100~M_{\rm Jupiter}$. The sine value of orbital inclination follows a uniform distribution from $1\fdg5$ to $11\fdg5$. The orbital eccentricity follows a beta distribution described in \citet{kipping13}. The argument of periastron follows a uniform distribution ranging from $0$ to $2\pi$ radian. The time of periastron passage follows uniform distribution which is determined by the orbital period. The stellar mass follows a Gaussian distribution using the measurements from \citet{monnier12}.  In each radial velocity sample, we scale Gaussian noise according to the uncertainties of the TRES measurements in \citet{hurt21}. We fit a flat line to each synthetic radial velocity curve using {\tt RadVal} \citep{fulton18}. 

For all simulated radial velocity samples, we consider a synthetic signal to be detectable in radial velocity, when its $p$-value $<0.001$ (i.e., $3.3\sigma$) while ignoring correlated noise \citep{hurt21}. At a specific radial separation from the star, we further require that a companion is detectable, when its mass is above the detection limit in Fig.~\ref{fig-contrast}. 

\begin{figure}[htb!]
\centering
	\includegraphics[width=0.475\textwidth]{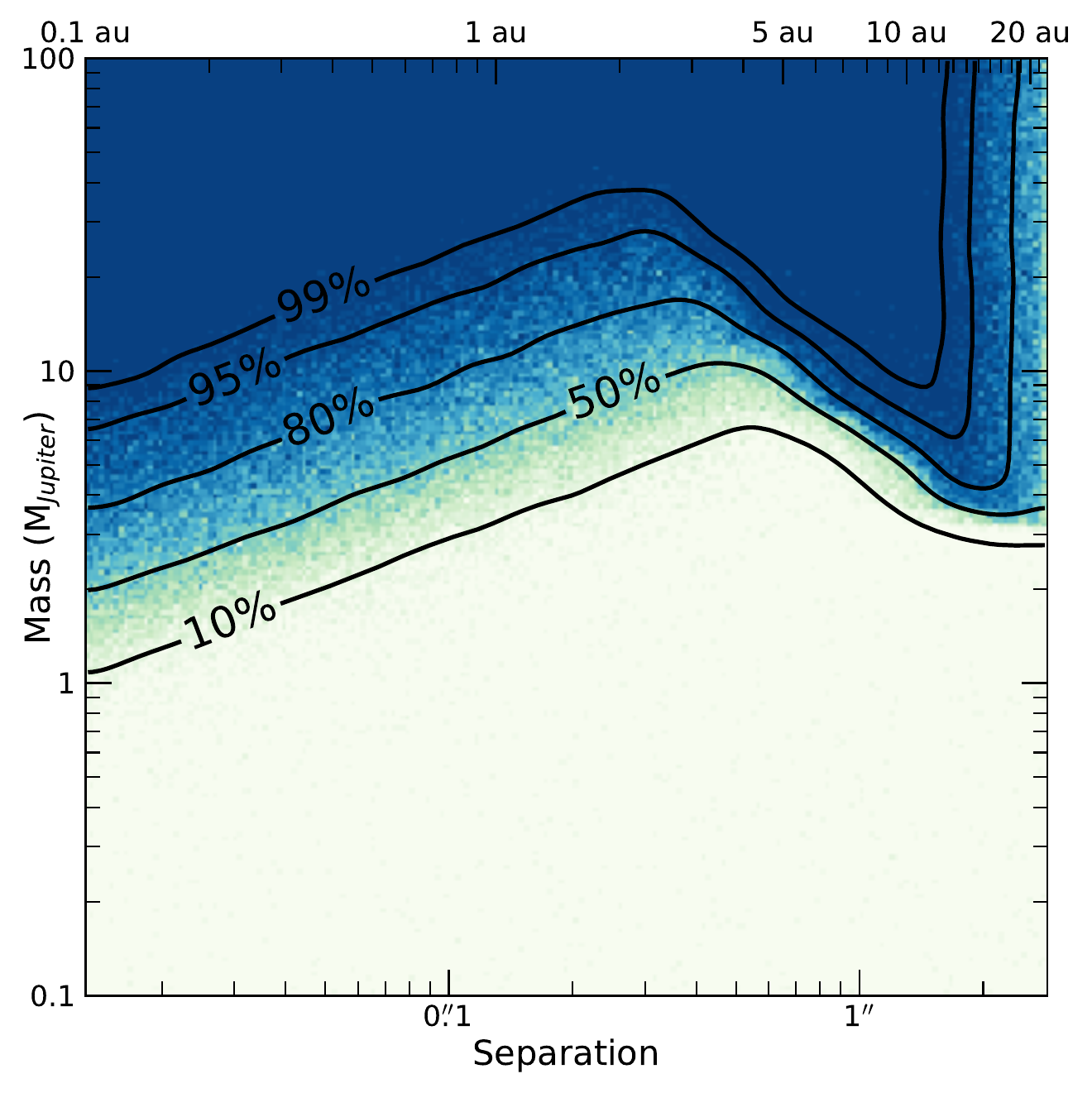}
    \caption{Detection probability of point sources as a function of semi-major axis of points sources for Vega, using a combination of the 2018 NIRC2 observation and the TRES radial velocity measurements in \citet{hurt21}, see Sect.~\ref{sec-comb-nirc2-tres}. We can reach ${\sim}8~M_{\rm Jupiter}$ detection limit at ${\sim}4$~au, while probing down to ${\sim}2$--${\sim}3~M_{\rm Jupiter}$ at $0.1$~au and $22$~au, respectively.}     \label{fig-limit-combined}    
\end{figure}

We present the detection probability of companions from the above injection-recovery procedure in Fig.~\ref{fig-limit-combined}. The detectability of companions from combined radial velocity and direct imaging follows two trends as a function of stellocentric separation. Interior to ${\sim}4$~au, the detectable planets increases with radial separation, approaching planets with ${\sim}8~M_{\rm Jupiter}$ down to ${\sim}1~M_{\rm Jupiter}$ at ${\sim}0.1$~au. Exterior to ${\sim}4$~au, the detectability is dominated by NIRC2 imaging, reaching down to ${\sim}2~M_{\rm Jupiter}$ at ${\sim}22$~au. Under the \citet{hurt21} framework, there is a possible non-absolute detection of companions near ${\sim}22$~au, which is limited by the NIRC2 field of view in our study, since the sampled orbital eccentricity adopted from \citet{kipping13} can position planets with semi-major axis less than ${\sim}22$~au beyond the $22$~au angular radius.

\subsection{Disk formation from mass limits}
The ALMA observation in \citet{matra20} resolves the outer dust belt of Vega extending from ${\sim}60$~au beyond ${\sim}150$~au. To explain the observed planetary system architecture, the authors have analyzed two scenarios that involves planets: either a chain of small planets within $70$~au with mass ${\gtrsim}6M_\oplus$, or a sole ${\sim}5~M_{\rm Jupiter}$ at $50$ -- $60$ au.

The combined Keck/NIRC2 $M_s$-band and TRES result can exclude the existence of ${\sim}8~M_{\rm Jupiter}$ planets from $0.1$~au to $22$~au. Despite the fact that with an age of ${\sim}400$~Myr there is no clear brightness difference between the hot-start and cold-start models \citep[e.g.,][]{spiegel12}, the observations presented here cannot rule out the sole giant planet which resides at $50$ -- $60$~au in \citet{matra20} for Vega.

For future exploration of the proposed sole giant planet at $50$~au to $60$~au using Keck/NIRC2 in $M_s$-band, a field of view that is between $6\farcs5$ and $7\farcs8$ is needed. With a pixel scale of $9.942$~mas, this corresponds to a half-width between $654$~pixel and $785$~pixel, which exceeds the current $512$~pixel half-width of the narrow camera for NIRC2 and thus not achievable. Alternatively, NIRC2 offers sampling setups that have pixel sizes of $20$~mas or $40$~mas, yet these setups are not feasible for $M_s$-band imaging due to the corresponding increased sky background. Nevertheless, the planned upgrades of the NIRC2 detector electronics may permit faster readouts to enable $M_s$-band imaging for such purposes. 

To explore far separation imaging for Vega using existing confirguration of NIRC2, either offsetting the vortex center away from the center (e.g., the observation of HR~8799~b using the Gemini Planet imager in \citealp{wang18}) of the NIRC2 narrow camera, or executing observations without the vortex coronagraph, could enable a half-width of $785$~pixel to test the sole giant planet scenario. With this $1024\times1024$~pixel field of view, the minimum permitted readout time is $0.18$~s\footnote{\url{https://www2.keck.hawaii.edu/inst/nirc2/ObserversManual.html\#Section2.4}}, which could enable a nearly identical exposure sequence in the 2018 dataset presented here. Given that extra parallactic angle change is needed to cover the entire field of view for these setups, they are beyond the scope of this study. 

\subsection{Implications for JWST observations}\label{sec-jwst-sim}
At an age  of $445\pm13$~Myr for Vega \citep{yoon10}, giant planets with several Jupiter mass do not have clear brightness distinction between different formation models in $M$-band \citep[e.g., Fig.~7 of][]{spiegel12}. For planets with less than ${\sim}5~M_{\rm Jupiter}$, their brightness is expected to peak at $4~\mu$m to $6~\mu$m \citep[e.g., Figure 6 of][]{spiegel12}. 

To image planets that are brightest at these wavelengths, we have applied the Keck/NIRC2 $M_s$-band which operates at a central wavelength of $4.67~\mu$m with a bandpass of $0.24~\mu$m.\footnote{\url{https://www2.keck.hawaii.edu/inst/nirc2/filters.html}} In comparison, the NIRCam instrument onboard the \textit{JWST} can cover nearly half the expected brightest wavelengths with its F444W filter ($4~\mu$m to $5~\mu$m).\footnote{\url{https://jwst-docs.stsci.edu/jwst-near-infrared-camera/nircam-instrumentation/nircam-filters}} By reaching a multi-Jupiter mass limit for $10$~au to $20$~au within this study, and one Jupiter mass or better beyond $20$~au with NIRCam \citep[][Fig.~5 therein]{meshkat18}, we can combine Keck/NIRC2 and \textit{JWST}/NIRCam to reach the deepest planet detection limit to investigate the planetary architecture for the Vega system.

Using the on-sky \textit{JWST}/NIRCam F444W contrast curve from \citet{carter22} in the \textit{JWST} ERS-1386 program \citep{hinkley22}, we calculate the expected on-sky contrast for Vega with NIRCam F440W in Fig.~\ref{fig-contrast} for \textit{JWST} GTO-1193 observations as follows. Assuming an identical instrument performance, which is an optimistic estimation given that the \textit{JWST} GTO-1193 uses MASK430R since it has a larger inner working angle -- and thus lower throughput -- than MASK335R in the \textit{JWST} ERS-1386 observations in \citet{carter22}, we scale the exposure times with SUB320 subarray (i.e., Observations 35 \& 36) from the \textit{JWST} GTO-1193 observations and recalculate the contrast based on exposure time difference. We then convert the \textit{JWST}/NIRCam F444W contrast to point source mass following \citet{carter21} for Vega while adopting the same age and apparent magnitude as for NIRC2.

In terms of reaching nominal contrast, the Keck/NIRC2 $M_S$-band observations perform better than NIRCam F440W within $1\arcsec$ and reaching a similar quality in exterior regions. In terms of mass detection limits, the Keck/NIRC2 $M_S$-band observations performs systematically better than the NIRCam F444W under optimistic assumptions above. In addition, the degradation of contrast close to the transmission near the edges of the the coronagraphic masks\footnote{Fig.~3 of \url{https://jwst-docs.stsci.edu/jwst-near-infrared-camera/nircam-instrumentation/nircam-coronagraphic-occulting-masks-and-lyot-stops}} from MASK335R in \textit{JWST} ERS-1386 to MASK430R in \textit{JWST} GTO-1193 has been ignored. All three aspects demonstrate that the NIRC2 $M_S$-band observations presented here establish the deepest high-contrast imaging exploration of planetary companions for Vega in the probed regions. Nevertheless, we emphasize that NIRCam should be better than NIRC2 for most other systems, since the study presented here should be the best case scenario for ground-based $M_S$-band imaging due to the brightness of Vega; for other targets that are fainter than Vega, they will have less favorable contrasts due to the relative background levels on the ground from NIRC2.

\subsection{Planet detection towards $0.1$~arcsec}
To obtain a more general understanding of planetary existence, the detection of intermediate separation planets near ${\sim}0\farcs1$ will likely yield the most discovery of planets \citep[e.g.,][]{nielsen19, fulton21}. In fact, for the detection of far-separation planets, existing direct imaging surveys equipped with extreme adaptive optics systems have experienced a degradation of contrast close to the central sources \citep[e.g.,][]{nielsen19, vigan21, xie22}. In comparison, for the detection of close-in planets, existing radial velocity surveys have less completeness for long orbital period planets \citep[e.g.,][]{wittenmyer20, fulton21}. As a result, in the detection probability map of companions for Vega in Fig.~\ref{fig-limit-combined}, there is a clear gap for middling separations.

To fill the gap, the concept of the vortex fiber nuller (VFN; \citealp{ruane19, echeverri20}) is designed to detect and characterize planets near and within $0\farcs1$. Using nulling interferometry in the near-infrared, VFN suppresses on-axis starlight while retaining off-axis companion light with acceptable loss, which increases the signal-to-noise ratio for companions and thus better enables their detection and characterization. The VFN is already installed on the phase 2 development of the Keck Planet Imager and Characterizer \citep[KPIC;][Echeverri et al., in preparation]{jovanovic20} on the Keck Observatory, providing access to planets between $30$ and $90$~mas in $K$-band. Furthermore, the limitation of VFN in localizing the companions in their orbit  will be further resolved with the concept of photonic lantern nuller \citep[PLN; e.g.,][]{xin22}. We expect that the application of VFN and PLN will fill the planet detection and characterization gap between direct imaging and radial velocity. For Vega, they will better explore possible hidden planets towards ${\sim}0\farcs1$ in Fig.~\ref{fig-limit-combined}. Nevertheless, the limitation of VFN in $K$-band is that it is less sensitive to evolved and cooler giant planets due to the drop in planet brightness in shorter wavelengths.

\section{Summary}\label{sec-sum}
We report $1.8$~h of $M_s$-band imaging observations of Vega using the Keck/NIRC2 vortex coronagraph. Despite a non-detection of companions, we have pushed the mass detection limits from existing high-contrast imaging observations in \citet{meshkat18} by nearly an order of magnitude smaller, see Fig.~\ref{fig-contrast}. Combining the NIRC2 results with existing radial velocity study using TRES in \citet{hurt21}, we can confidently rule out companions more massive than ${\sim}8~M_{\rm Jupiter}$ from $0.1$~au to $22$~au for Vega. Within this range, NIRC2 can reach planets that are less massive than $5~M_{\rm Jupiter}$ beyond $9$~au.

While the NIRC2 observations presented here are sensitive to planets of $5~M_{\rm Jupiter}$ at $9$~au down to $3~M_{\rm Jupiter}$ at $22$~au, it is limited by the field of view and thus cannot be use to test the scenario in \citet{matra20} that one multi-Jupiter mass planet at $50$~au to $60$~au is responsible for the planetary system architecture. Alternatively, observing without using the vortex coronagraph while perform multi-point dithering\footnote{\url{https://www2.keck.hawaii.edu/inst/nirc2/ObserversManual.html\#Section3.2.10}}, combined with extra parallactic angle change, may provide enough sensitivity to image such a perturber for the Vega system by pushing deeper than \citet{heinze08} for these separations. 

Combining Keck/NIRC2 in $M_s$-band and NIRCam in its F444W filter assuming identical performance with \textit{JWST} from ERS-1386 observations \citep{hinkley22, carter22}, we can reach comparable detection limits of companions interior to $20$~au. Interior to ${\sim}10$~au, although companion imaging limits increases from ${\sim}5~M_{\rm Jupiter}$ at $10$~au to ${\sim}100~M_{\rm Jupiter}$ at $2$~au in Fig.~\ref{fig-contrast}, this region is where radial velocity measurements can provide detection limits of multi-Jupiter or better in \citet{hurt21}. Future works following \citet{mawet19} and \citet{Llop-Sayson21} in combining measurements including direct imaging here and radial velocity in \citet{hurt21}, as well as upcoming imaging observations with \textit{JWST} GTO-1193, in addition to the VFN and PLN concepts to detect planets towards ${\sim}0\farcs1$, can enable us to obtain the most holistic understanding for the planetary system of this historical photometric standard star -- Vega.

\begin{acknowledgements}
We thank the anonymous referee for their constructive comments that increased the clarity and reproducibility of this paper. This research is partially supported by NASA ROSES XRP, award 80NSSC19K0294. B.R.~has received funding from the European Research Council (ERC) under the European Union's Horizon 2020 research and innovation programme (PROTOPLANETS, grant agreement No. 101002188). \'E.C.~has received funding from the European Research Council (ERC) under the European Union's Horizon Europe research and innovation programme (ESCAPE, grant agreement No 101044152). Some of the data presented herein were obtained at the W.~M.~Keck Observatory, which is operated as a scientific partnership among the California Institute of Technology, the University of California and the National Aeronautics and Space Administration. The Observatory was made possible by the generous financial support of the W.~M.~Keck Foundation. The authors wish to recognize and acknowledge the very significant cultural role and reverence that the summit of Maunakea has always had within the indigenous Hawaiian community.  We are most fortunate to have the opportunity to conduct observations from this mountain. Part of the computations presented here were conducted in the Resnick High Performance Computing Center, a facility supported by Resnick Sustainability Institute at the California Institute of Technology.
\end{acknowledgements}

\bibliography{refs}
\end{CJK*}
\end{document}